\documentclass[aip,10pt,floatfix,twocolumn]{revtex4-1} 
\usepackage{blindtext}
\usepackage{graphicx}
\usepackage{graphics}
\usepackage{bm}   % for math
\usepackage{verbatim}   % for math
\usepackage{amsfonts}
\usepackage{amsmath}
\usepackage{bm}
\usepackage{amssymb}
\usepackage{subfigure}
\usepackage{adjustbox}
\usepackage{lipsum}
\usepackage{xcolor}

\usepackage{adjustbox}

\usepackage{stfloats}

\usepackage{amsmath,amscd}
\usepackage {extarrows}
\usepackage{algorithm,algpseudocode}
	
\makeindex

\begin{document}

\title[]{Athermal quasistatic cavitation in amorphous solids: effect of random pinning}
% Force line breaks with \\
\author{Umang A. Dattani}
\email{umangad@imsc.res.in}
\affiliation{The Institute of Mathematical Sciences, C.I.T. Campus,
Taramani, Chennai 600113, India}
\affiliation{Homi Bhabha National Institute, Anushakti Nagar, Mumbai, 400094, India}

\author{Smarajit Karmakar}%
 \email{smarajit@tifrh.res.in}
\affiliation{Tata Institute of Fundamental Research, 36/P, Gopanpally Village, Serilingampally Mandal,Ranga Reddy District, 
Hyderabad, 500046, Telangana, India}
%}%

\author{Pinaki Chaudhuri}
 \email{pinakic@imsc.res.in}
\affiliation{The Institute of Mathematical Sciences, C.I.T. Campus,
Taramani, Chennai 600113, India}
\affiliation{Homi Bhabha National Institute, Anushakti Nagar, Mumbai, 400094, India}

\date{\today}% It is always \today, today,

\begin{abstract}
Amorphous solids are known to fail catastrophically via fracture, wherein cavitation at nano-metric scales is known to play a significant role.
Micro-alloying via inclusions is often used as a means to increase the fracture toughness of amorphous solids. Modeling such inclusions as randomly pinned particles that move only affinely and do not participate in plastic relaxation, we study how the pinning influences the process of  cavitation-driven fracture in an amorphous solid. Using extensive numerical simulations and probing in the athermal quasistatic limit, we show that  just by pinning a very small fraction of particles, the tensile strength is increased and also the cavitation is delayed. Further, the cavitation that is expected to be spatially heterogeneous becomes spatially homogeneous by forming a large number of small cavities instead of a dominant cavity. 
\end{abstract}

\maketitle

\section{Introduction}

The mechanical properties of amorphous solids are utilized in diverse applications in industries and our daily lives \cite{schuh2007mechanical, bonn2017yield}. Therefore, mechanical failure of these materials is an area of concern. Hence, understanding the physical processes that lead to the failure of these structurally disordered solids is a domain of current active research, with the primary goal being to figure out design pathways that can sustain against such failures \cite{rodney2011modeling, liu2021elastoplastic}. Cavitation, i.e., the formation of nano-cavities within the solid, has been identified as a precursor to eventual failure via fracture \cite{bouchaud2008fracture, murali2011atomic}. In experiments, the fracture in amorphous solids has been shown to propagate via the coalescence of cavities in the solid along the direction of the crack \cite{shen2021observation}, which has motivated numerical investigations to analyze the mechanisms underlying the cavitation process \cite{Falk, Pinaki, Pablo, Pablo2, CavitySmarajit}. Recently \cite{FirstWork}, it has been demonstrated that the plasticity associated with cavitation has the same universal characteristic elastoplastic response of amorphous solids undergoing failure via cavitation under uniform expansion\cite{FirstWork} and on exploring a combination of loading scenarios where cavitation can occur\cite{SecondWork}. In Ref.\cite{SecondWork}, we demonstrated that a combination of deformation processes, for example, uniform expansion followed by oscillatory shear of a certain amplitude, can enhance the formation of cavities at densities that are much higher than the cavitation density observed in uniform expansion only. These results suggest that in a natural deformation process in which various forms of deformations will be coupled together, the material can show unpredictable failure behavior, which will be difficult to control. Thus a systematic study of how such a cavity-dominated failure process in an amorphous solid can be mitigated effectively will be of significant interest for practical applications.
%It thus becomes important to understand the role of these cavitation instabilities in amorphous solids and how they can be prevented. 

In the quest of making glasses with higher fracture-toughness, the seeding of the amorphous solids with micro-alloyed inclusions has gained a lot of popularity in the last few years \cite{PinningNature,PinningIntro,torquato2002random, tyukodi2016finite, garrett2012effect, gonzalez2016role}. In the numerical modeling, a minimal model of these micro-alloyed inclusions describes these inclusions as pinned/frozen particles \cite{ItamarPinning,Bhanu,Bhanu2,liu2023pinningLengthScale}. In the context of probing their mechanical behavior, it is assumed that they only move affinely during the deformation and do not actually undergo non-affine motion. Using such a model system, systematic studies investigating the response of a pinned amorphous solid to a shear-deformation have explored the microscopic theories\cite{ItamarPinning}, yielding mechanisms\cite{Bhanu}, suppression of shear-banding ~\cite{Bhanu2}, development of intrinsic length-scales ~\cite{Bhanu,liu2023pinningLengthScale} etc. As these studies focus on the shear-deformation of high-density amorphous solids, they do not access the region where cavitation instabilities occur, i.e., under axial tension \cite{Pablo,FirstWork,SecondWork}. Due to the importance of cavitation instabilities in fracture, it becomes important to study the response of an amorphous solid under deformation modes where cavitation can occur. 

In this work, we, therefore, study the response of amorphous solids with micro-alloyed inclusions modeled as pinned particles under uniform expansive deformation under athermal quasistaic conditions. We find that, even in presence of a very small fraction of pinned particles, cavitation becomes more spatially homogeneous and the cavitation point shift to lower densities and lower pressures, implying a higher load-bearing capacity of the pinned solid. The sharp brittle-yielding-like transition seen in unpinned solids becomes more gradual with significantly smaller sizes of plastic events due to which system size dependence becomes very weak. On tracking the eigenvalues of the Hessian near cavitation instabilities, we find that, on the potential energy landscape, cavitation occurs via a saddle-node bifurcation and the average spatial decay of displacements in the plastic eigenmodes from the plastic center reveal a length-scale in pinned solids that parallels the length-scale set by average distance between two pinned sites. The presence of a length-scale explains the absence of system-size effects and the drastic decrease in the mean sizes of plastic events. Our findings thus reveal how micro-alloyed inclusions can suppress the cavitation and how the presence of a length-scale of plasticity controls the deformation response of such a pinned solid.  

%\newpage

The manuscript is organised as follows. After initial introductory discussion in Section I, we provide in Section II a brief overview of the model amorphous solid that we consider for our study and the methodology of our simulations as well as analysis. In Section III, we discuss the detailed findings of our investigations regarding the presence of random pinning and its influence on the cavitation process. Finally, we provide a concluding discussion in Section IV.

\section{Model and Methods}

\subsection{Model details and initial states}

{We use the well-characterized two-dimensional model consisting of two species labelled $A$ and $B$ at $65:35$ concentration ratio, interacting via pairwise Lennard-Jones potential. The interaction parameters are -- $\sigma_{AA}=1.0$, $\sigma_{BB}=0.88$, $\sigma_{AB}=0.8$, $\epsilon_{AA}=1.0$, $\epsilon_{BB}=0.5$, $\epsilon_{AB}=1.5$ \cite{2DKA}. With this model, we smoothen the interaction potential up to first two derivatives. The form of the interactions between $i^{th}$ and $j^{th}$ particle becomes:
\begin{equation}
    \phi(r_{ij})=4\epsilon_{\alpha\beta}\left[\left( \frac{\sigma_{\alpha\beta}}{r_{ij}} \right)^{12} - \left(\frac{\sigma_{\alpha\beta}}{r_{ij}} \right)^6\right] + u(r_{ij})
\end{equation}
where,
\begin{equation}
u(r_{ij})=C_{0} + C_{2}\left(\frac{r_{ij}}{\sigma_{\alpha\beta}}\right)^2 + C_{4}\left(\frac{r_{ij}}{\sigma_{\alpha\beta}}\right)^4
\end{equation}
Here, $\alpha$ and $\beta$ correspond to either of the labels $A$ or $B$.
The constants $C_0$, $C_2$ and $C_4$ are determined by requiring the potential and its first two derivatives to be zero at the cutoff $r=2.5\sigma_{ij}$. The simulations have been performed for a variety of system sizes ranging from $N=10^3$ to $N=10^5$.}
\subsection{Initial states}
{To prepare initial states for our study, we first equilibriate the system at $T=1.0$ (in LJ units), which is in the liquid regime, followed by cooling at a constant rate of $10^{-4}$ per MD timestep to a final temperature of $T=0.01$ \cite{Bhanu}, which is in the glassy regime. The corresponding glass transition temperature of the model system is at $T=0.44$ \cite{2DKA}. The athermal states used in our study are generated by obtaining inherent structure states corresponding to the glassy configurations at $T=0.01$, via conjugate gradient (CG) minimization\cite{cg}.}

\subsection{Athermal Quasistatic Expansion}

{Starting from a spatially homogeneous high density state ($\rho=1.2$ for KABLJ) having positive barostatic pressure, we study the athermal quasi-static response  (i.e. in the absence of any thermal effects and in the limit of vanishing driving rates) of this system to isotropic expansion \cite{Pablo}. In each expansion step, a constant volume strain is applied on the system by rescaling the length of the box by a factor $(1+\epsilon)$ along with affine transformation of particle coordinates, followed by minimization of the energy of this strained configuration using the conjugate gradient algorithm~\cite{cg}.  The values of $\epsilon$ are varied from $\epsilon=10^{-4}$ to $\epsilon=10^{-9}$. The AQE simulations are done using LAMMPS\cite{LAMMPS}. }

\subsection{Pinning}
We choose a small fraction of particles $c=0.01$ to $c=0.05$ in the generated solid and freeze their motion. The particles are chosen randomly as long as no two pinned particles lie within the cut off of the interaction between them\cite{Bhanu}. This helps avoiding the scenario where two close-by pinned sites increase the energy of the system. The pinned particles only move affinely when the strain is applied. During the energy minimization, these pinned particles are not allowed to move.

\subsection{Hessian of potential energy}
{LAPACKE\cite{lapack99} is used for doing the stability analysis of the local minima states, by computing eigenvalues and eigenvectors of the Hessian matrix $\mathcal{H}^{\alpha \beta}_{i j}$, which is defined as 
\begin{equation}\label{hessian}
    \mathcal{H}_{ij}^{\alpha \beta} = \frac{\partial^{2} U
    \left(\{\mathbf{r}_{i}\}\right)}{\partial r_{i}^{\alpha} \partial r_{j}^{\beta}},
\end{equation}}
where $U\left(\{\mathbf{r}_{i}\}\right)$ is the potential energy of the system and $\mathbf{r}_i$ is the position vector of particle $i$. The indices $\alpha, \beta \in \{x,y\}$ whereas $i,j \in \{1, \ldots , N\}$. %Our studies are done for a variety of system sizes ranging from $N=10^3$ to $N=10^5$.

{If we now consider a system of $N$ particles, where particle numbers, $i=1,\cdots, m$ are free and particle numbers $i=m+1, \cdots ,N$ are pinned, then the potential energy of such a system can be expressed as 
\begin{equation}\label{equation1}
 U(r)=\frac{1}{2}\left[\sum_{i,j=1; i\neq j}^{m}\phi_{ij} + 2\cdot \sum_{i=1}^m \sum_{j=m+1}^{N} \phi_{ij} \right], 
 \end{equation}
where the first term comes from the interactions between unpinned particles, the second term comes due to the interactions between the pinned and the unpinned particles. Note that the term due to interactions between pinned sites is set to zero because of our pinning protocol}

{By substituting Eq.(\ref{equation1}) in Eq.(\ref{hessian}) , it can be shown that the first term in the sum of Eq.(\ref{equation1}) gives a contribution, 
\begin{equation}\small
       H^{ij}_{\alpha\beta}=-\sum_{k,i; k\neq i}  \left[ \left( \frac{\phi_r}{(r^{ik})^3} - \frac{\phi_{rr} }{(r^{ki})^2}  \right)r^{ki}_{\alpha} r^{ki}_{\beta}
       - \delta_{\alpha\beta} \frac{\phi_r}{r^{ki}} \right](\delta^{ji}-\delta^{jk}). 
\end{equation}
The second term of the sum in Eq.(\ref{equation1}) gives,
\begin{equation}
    H^{ij}_{\alpha\beta}=-2\, \sum_{k=0}^{m}\,\sum_{l=m+1}^{N} \left[  \frac{\phi_{r}\,r^{kl}_{\alpha}\,r^{kl}_{\beta}}{(r^{kl})^3} - \frac{\phi_{rr}\,r^{kl}_{\alpha}\,r^{kl}_{\beta}}{(r^{kl})^2} - \delta_{\alpha\beta} \frac{\phi_r}{r^{kl}}   \right]  \delta^{ij},
\end{equation}

}
{where $\phi_r$ and $\phi_{rr}$ are first and second derivatives of the pair-potential with respect to variable $r$ respectively.}

\begin{figure*}[]
    \includegraphics[width=0.99\linewidth]{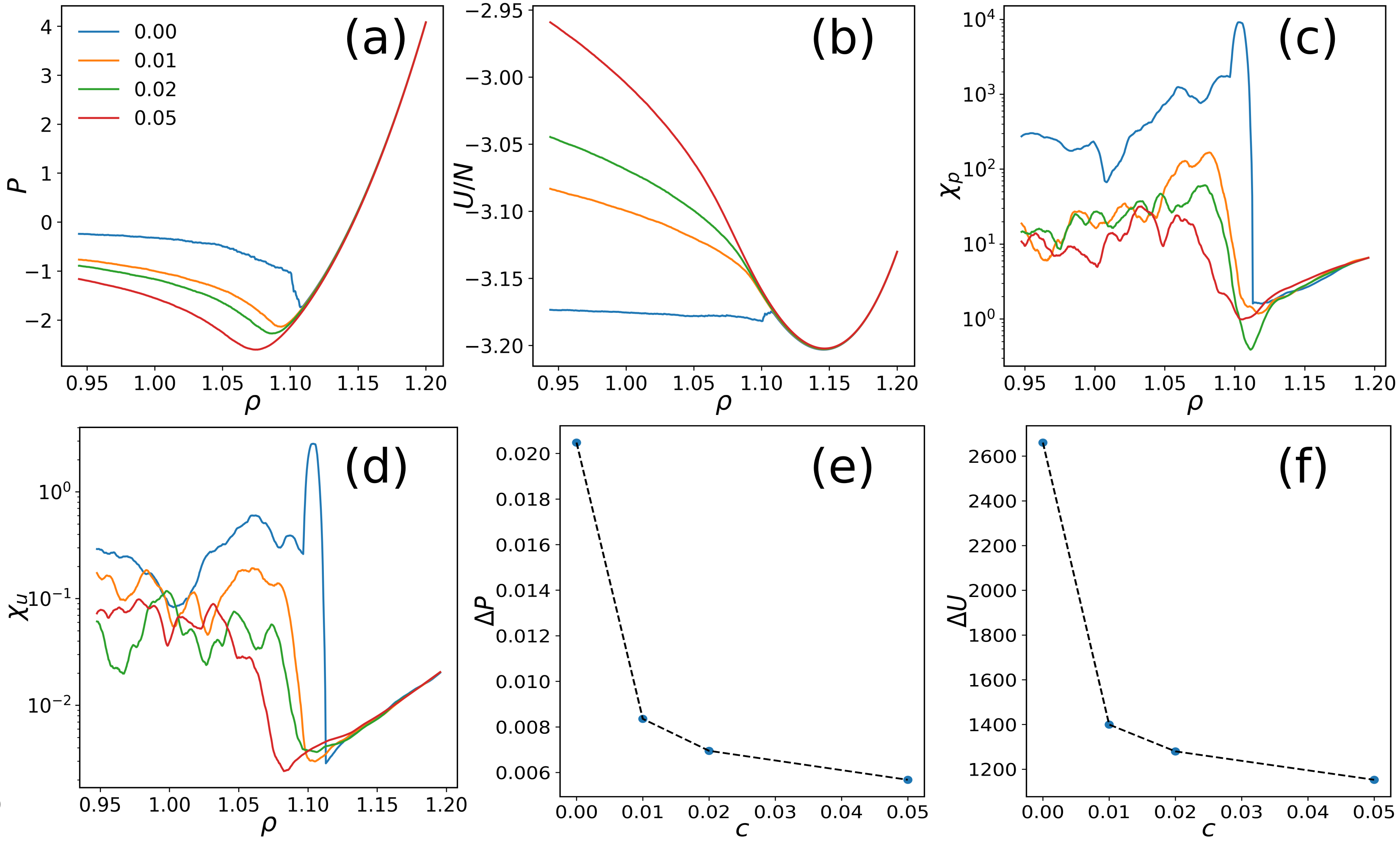}
    \caption{For $N={10}^5$ and different pinning concentrations, $c$, as marked: Variation with density $\rho$ of (a) Pressure $P$ (b) Energy per particle $U/N$ (c) Pressure susceptibility $\chi_P$ (d) Energy susceptibility $\chi_U$. Change in (e) average size of pressure jump $\Delta{P}$, and (f) average size of energy drops $\Delta{U}$, with pinning concentration.  }
    \label{PressDens}
\end{figure*}

\begin{figure*}[]
    \centering
    \includegraphics[width=0.99\linewidth]{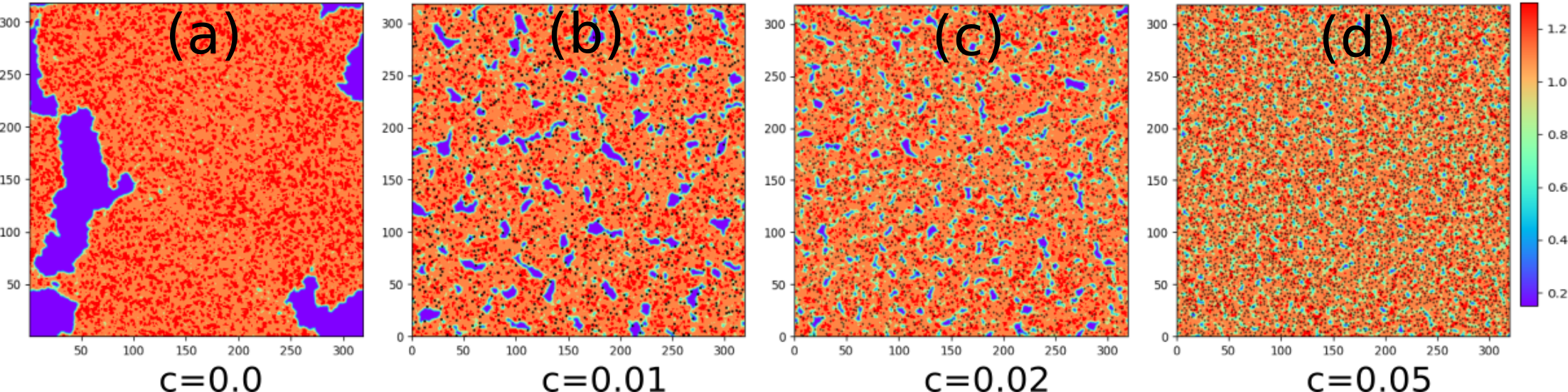}
    \caption{Density field for different pinning concentrations, $c=$ 0.00 (a), 0.01 (b), 0.02 (c), 0.05 (d), measured at $\rho=0.982$.}
    \label{DensityMaps}
\end{figure*}

\begin{figure*}[]
   \includegraphics[width=0.99\linewidth]{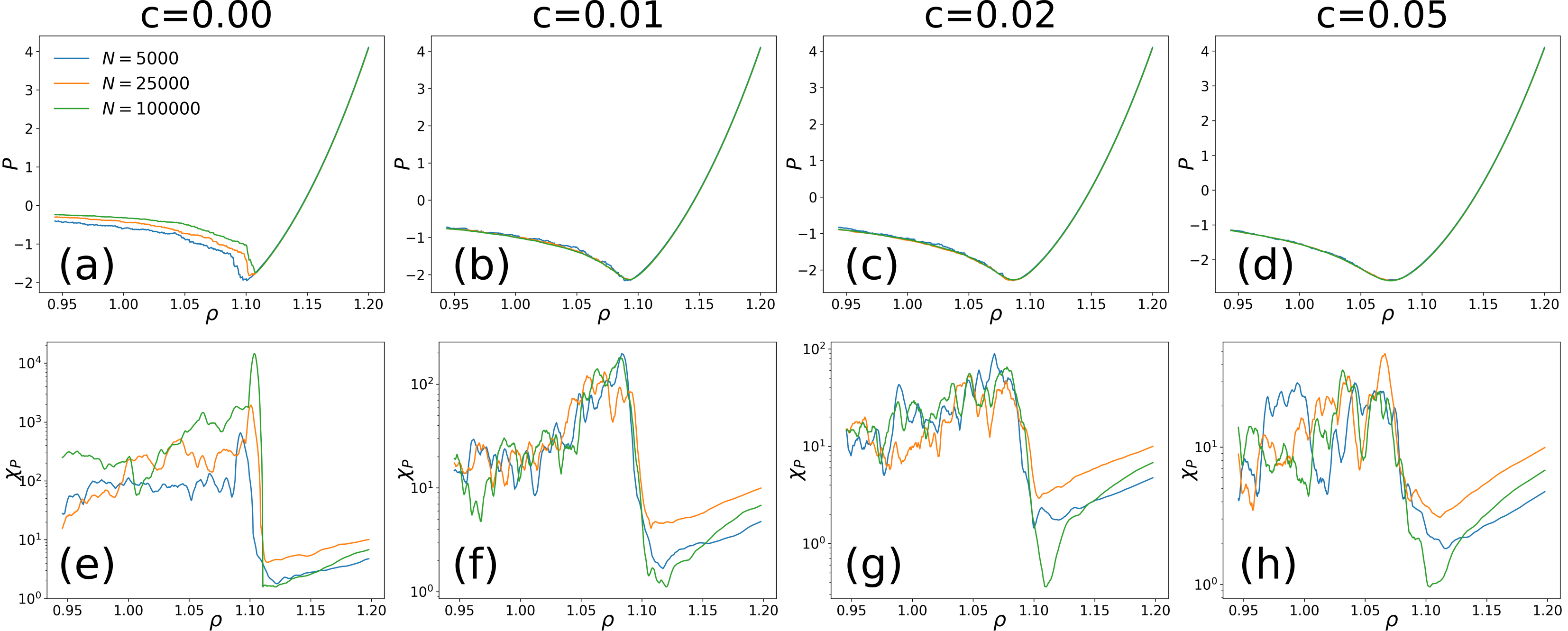}
  \caption{(Top) Variation of ensemble averaged pressure ($P$) with density ($\rho$) for different system sizes, as labelled, at $c=$ 0.00 (a), 0.01 (b), 0.02 (c), 0.05. (d) (Bottom) Corresponding variation of pressure susceptibility $\chi_{p}$ with density $\rho$, in each case.}
  \label{systemSize}
\end{figure*}

\begin{figure*}[]
    \centering
    \includegraphics[width=0.99\linewidth]{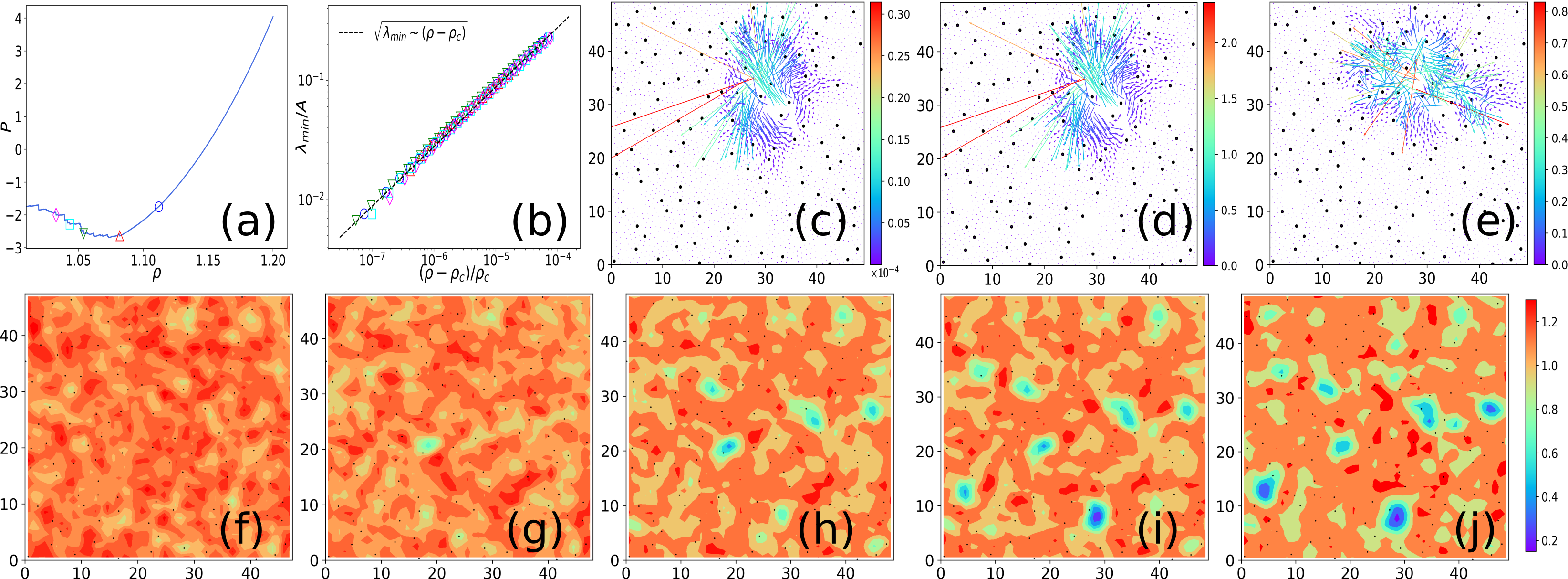}
    \caption{(a) Pressure $P$ vs density $\rho$ for an expansion trajectory, using $N=2500$, corresponding to $c=0.05$. (b) Demonstration of square-root singularity for lowest eignevalue of the Hessian, $\lambda_{\rm min}$ (appropriately scaled by fit parameter $A$), computed at the density points in (a) corresponding to occurrence of plastic instabilities; $\rho_c$ is the estimate density location of the event in each case. (c) Eigenmode, (d) displacement field just before the drop and (e) displacement fields across the plastic drop, occurring at one such plastic event, near $\rho \approx 1.03277$. (f)-(j) Evolution of the density field across the expansion trajectory shown in (a). }
    \label{squareroot}
\end{figure*}

\section{Results}

\subsection{Yielding and spatial ramifications}
Upon expanding the amorphous solid isotropically under quasistatic loading, as discussed in previous works\cite{Pablo,FirstWork}, the pressure of the solid decreases monotonically, eventually reaching negative values. After a certain threshold, a sharp jump in the pressure accompanies the cavitation of the solid. Upon further expansion, the cavities grow and merge, leading to system-spanning fracture of the solid~\cite{FirstWork}. Here, we expand the same amorphous solid isotropically but with a small fraction of particles pinned(frozen) and compare it with the case without pinning\cite{Pablo,FirstWork}. As shown in Fig.~\ref{PressDens}(a), compared to the unpinned solid, the pinned solid, on average, does not show a large pressure jump that is seen around the first cavitation event. The location of the yield point, which is usually marked by a turn in the pressure-density curves, shifts to lower and lower densities with increasing concentration of pinned particles, $c$, demonstrating an increase in the load-bearing capacity of pinned solids. In the pinned solids ($c\neq 0$), the turning of the curve occurs due to small pressure jumps that occur gradually, leading to a smooth average pressure vs density ($P - \rho$) curve as opposed to the ensemble-averaged trajectories of an unpinned solid ($c = 0$), which show an abrupt jump. A similar trend reflects in the per particle energy vs density ($U/N - \rho$) plots in Fig.~\ref{PressDens}(b) i.e. no large energy drop shows up in the pinned solid unlike the case of unpinned solid and in fact, the energy per particle of the pinned solid keeps increasing with increasing pinning concentration. The nature of critical-like behaviour in the ($P - \rho$) or ($U/N - \rho$) plots can be studied by measuring the fluctuations of pressure and energy at a given value of density across ensembles as characterized by the following susceptibilities\cite{FirstWork}, 
\begin{eqnarray}
    \chi_p(\rho)&=&N\,\left( \langle P^2(\rho)\rangle - \langle P(\rho)\rangle^2 \right), \quad \mbox{and}\\
    \chi_u(\rho)&=&(1/N)\,\left( \langle U^2(\rho)\rangle - \langle U(\rho)\rangle^2 \right)
\end{eqnarray}
  and shown in Fig.~\ref{PressDens}(c) \& (d) respectively. A sharp susceptibility peak around the big pressure jump / energy drop seen in the unpinned solid~\cite{FirstWork} is tamed down due to pinning. For higher pinning concentrations, a clear peak around the yielding is not seen implying that the yielding of the solid is more localised and gradual. The gradual nature of cavitation in pinned solids is also echoed in the average size of pressure jumps $\langle\Delta P\rangle$ and energy drops $\langle\Delta U\rangle$ encountered during expansion (see \ref{PressDens}(e) \& (f)) for different pinning concentrations $c$. The average size of avalanches, $\langle\Delta P\rangle$ \&  $\langle\Delta U\rangle$ decreases drastically with increasing values of $c$ implying suppression of large pressure jumps and energy drops. 
  
To probe the spatial-nature of cavitation in pinned solids, we look at the coarse-grained spatial density maps for different values of $c$ at a same value of density ( shown in Fig.~\ref{DensityMaps}). They suggest that the cavitation occurs at multiple sites in the solid gradually over the course of expansion for $c\neq 0$ instead of a heterogeneous cavitation starting with a big cavity for $c = 0$. The creation of a large number of cavities in the system also explains the increase in energy with expansion seen in Fig.~\ref{PressDens}(b) as presence of a large number of particles on the surfaces is expected to increase the energy of pinned solids. So, to summarise, pinning smooths out the sharp brittle yielding-like cavitation transition seen in amorphous solids without any micro-alloying and causes less heterogeneous cavitation. This scenario is consistent with a previous study on effect of pinning on yielding transition under simple shear\cite{Bhanu} where pinning made the yielding more spatially homogeneous.

\subsection{System-size effects}
In Ref.~\cite{FirstWork}, strong system size effects were observed in the $P - \rho$ the $\chi_p$ curves for athermal quasistatic expansion of amorphous solid. Hence it is important to probe the dependence on system sizes for pinned solids as well. The $P - \rho$ plots and $\chi_p - \rho$ plots for different system sizes across the pinning concentrations are shown in Fig.\ref{systemSize} top panel and bottom panels, respectively. Unlike the case of unpinned solids, the $P-\rho$ and $\chi_p-\rho$ curves for pinned solids show little to no dependence on the system size. This occurs due to the emergence of an intrinsic length scale of plasticity $\xi<<L$ in these systems due to the imposed random pinning constraints, which are discussed in the subsequent paragraphs.

%\begin{figure*}
 %   \centering
  %  \includegraphics[width=0.95\linewidth]{Figures/Density_field.png}
   % \caption{Coarse-grained density field of the solid for different pinning concentrations at same density shows contrasting morphologies of cavities in the solid. Black dots represent pinned sites in the solid}
    %\label{DensityMaps}
%\end{figure*}

\subsection{Irreversible plastic events on the potential energy landscape}
Under athermal quasistatic shear, near a plastic instability, the lowest non-zero eigenvalue (apart from the two zero-modes in 2D) of the hessian matrix of the potential energy Eq.~\ref{hessian} is known to vanish as a square root of strain difference from the point of instability ~\cite{MalandroLacks,MaloneyLemaitre}. i.e. $\lambda_{min}\sim\sqrt{\gamma_c -\gamma }$, where $\lambda_{min}$ refers to the minimum eigenvalue. This occurs due to a saddle-node bifurcation on a potential energy landscape (PEL) where the local minima in which the system resides becomes unstable in one direction. One of the possible ways to arrive at such a square-root power law has been discussed in Ref.~\cite{SquareRootDerivation}, where it has been shown how the nature of non-affine displacements in amorphous solids near a plastic instability with only $\lambda_{min}\to 0$ gives rise to the square-root singularity. Vanishing of only the lowest non-zero eigenvalue near the plastic instability ensures that the spatial map of eigenmode corresponding to the vanishing eigenvalue dominates the displacement field on the approach to such a plastic instability ~\cite{MaloneyLemaitre}. The same scenario of a square-root singularity with only one vanishing eigenvalue was shown to hold under athermal quasistatic expansion on approach to a cavitation instability as well \cite{FirstWork}. 

In the current context of the athermal quasistatic expansion of pinned solids, we find that the same square-root singularity scenario holds. In Fig.~\ref{squareroot}(a), we show one such trajectory corresponding to $c=0.05$ \& $N=2500$. Fig.\ref{squareroot}(b) shows the square root singularity at which the lowest-eigenvalue of the Hessian vanishes as $\lambda_{min}\sim\sqrt{(\rho-\rho_c)/\rho_c}$ on approach to the plastic instability at the points marked in Fig.\ref{squareroot}(a), where $\rho_c$ is the point at which the plastic instability occurs. The Fig. \ref{squareroot}(d)-(e) show an eigenmode on the approach to the plastic instability, displacement field on the approach to the plastic instability, and the displacement field across the pressure drop, respectively, for one of the pressure-jumps in Fig.\ref{squareroot}(a)\& (b). As evident from the vector-field maps, the displacement fields on the approach to instability are predicted by the eigenvector of $\lambda_{min}$, but the displacement fields across the pressure jump do not have a high overlap with the eigenmode on the approach to the instability. This occurs because of the cascade/avalanche nature of the plastic jump. These avalanches are also more localized spatially due to pinning, unlike those seen in unpinned solids which can be system spanning\cite{AValanches1,Avalanches2,FirstWork}. 
 All in all, the mechanism of plasticity on PEL being analogous to unpinned solids does not come as a surprise because pinning only blocks certain pathways of relaxation on the PEL of unpinned solids ~\cite{WalesEnergyLandscape}.

\begin{figure*}[ht]
    \centering
    \includegraphics[width=0.9\linewidth]{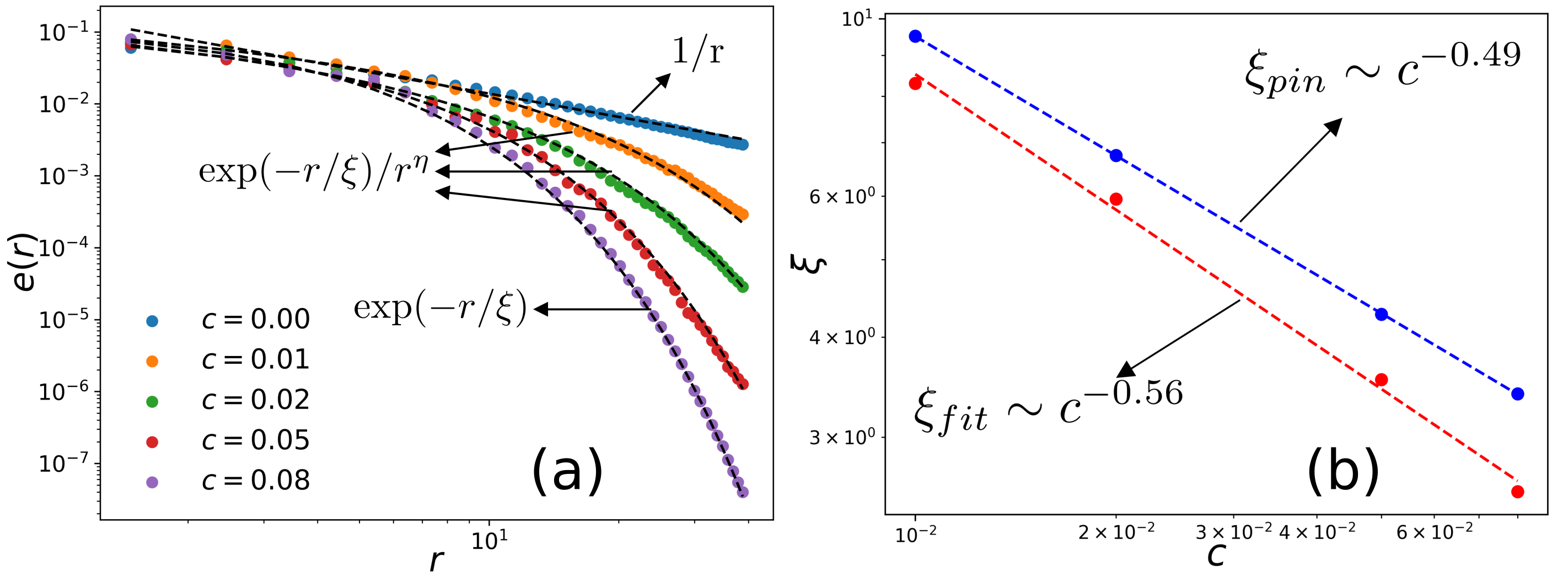}
    \caption{(a) Spatial decay profile of magnitude of displacements in the eigenmode before the yielding, $e(r)$, leading to the plastic events, shown for $c=0.00, 0.01, 0.02, 0.05, 0.08$, with averaging done over $10$ plastic events for each case. $r$ is the distance from the plastic center defined by a particle undergoing maximum displacement in a mode. (b)Dependence of length scale ($\xi_{\rm fit}$) extracted from fits to $e(r)$ (see text for discussion) with pinning concentration; also shown is the behaviour of $\xi_{\rm pin}$, extracted from pinning density, with $c$.}
    \label{decayEigenmodes}
\end{figure*}

\subsection{Spatial decay of plastic modes and a length scale of plasticity}

The plasticity under shear deformation is known to occur via localised rearrangements of particles in a shear transformation zone ~\cite{Argon1979STZ,FalkLangerSTZ,LernerSTZ}. These localised rearrangements(and their eigenmode) have a quadrupolar shape \cite{eshelby1957determination,LukaLernerPlasticModes} and the radial part of the displacements of the medium decay as  $r^{-(d-1)}$;  where $r$ is the distance from the center of rearrangement, and $d$ is the spatial dimensions. Even though the rearrangements are local, the displacement fields have a long-range character, i.e., in 2d, they decay as $1/r$ from the center. The eigenmodes, $\vec{e}$, corresponding to the vanishing eigenvalue $\lambda_{min}$ are known to decide the direction of failure on the PEL for both shear and expansion ~\cite{MaloneyLemaitre,EigenmodesTanguy,SoftModesManningLiu,FirstWork}. 

In the context of pinning, we, too, look at the radial decay profile of displacements from the plastic center ~\footnote{defined as particle undergoing highest displacement in an eigenmode} of the eigenmode $\vec{e}$, just before the plastic event, i.e., at a distance of $\delta\rho\approx 10^{-7}$, in the pre-yield regime.
%and compare it with unpinned cases. 
We choose to study the spatial profiles in the pre-yield regime,  where density inhomogeneities (which can interfere with the profile shapes) are largely absent. The decay profiles averaged over $10$ plastic events for each pinning concentration are shown in Fig.\ref{decayEigenmodes}(a). For the unpinned case, as one would expect, the decay profile $e(r)\sim 1/r$. For the pinned solids, interestingly, we find two regimes. For the smaller pinning concentrations, \textit{viz}. $c=0.01$, $c=0.02$ and $c=0.05$,  we see that the decay profile fits a screened power-law function $e(r) \sim \exp(-r/\xi)/r^{\eta} $ with $\eta\approx 0.42$ for $c=0.01$ and $c=0.02$ whereas $\eta=0.13$ for $c=0.05$. For larger pinning concentration, \textit{viz}. $c=0.08$,  we find that the decay profile is well-fitted by an exponential form $e(r)\sim \exp(-r/\xi)$ suggesting a crossover from a screening-like decay to a purely exponential decay. The scale-dependent exponential part of the decay profile allows us to extract a length scale $\xi$ from the fit parameters. The values of $\xi_{fit}$ for different pinning concentrations are shown using red color in Fig.\ref{decayEigenmodes}(b).

For a fixed pinning concentration $c$, one naturally obtains a length scale $\xi_{pin}=\sqrt{1/(c\cdot\rho)}$, which denotes an average distance between any two pinned sites. But, since the eigenmode decay profiles shown in Fig.\ref{decayEigenmodes}(a) are sampled at different densities from the expansion trajectory, the blue curve showing $\xi_{pin}=\sqrt{1/(c\cdot\rho)}$ for the sampled densities in Fig.\ref{decayEigenmodes}(b). The length scale extracted from the fits in Fig. \ref{decayEigenmodes}(a) thus shows reasonable parallels with the length scale introduced by pinning as seen in Fig.\ref{decayEigenmodes}(b). The decay exponents of length scales with pinning concentrations for $\xi_{pin}$ is slightly different from exponent $0.5$ primarily due to the changes in density where these decay profiles are obtained for each pinning concentration.

%SK: Umang, please clarify this. Umang: Done!

The exponential nature of the decay profiles and the extracted length scales thus suggest that, with increasing pinning concentrations, the displacements in the eigenmode become more and more localized. This implies that the consequence of plastic events in a pinned solid has reduced spatial effects, unlike the unpinned solid, which has a non-local power-law character.  Hence, this will restrict the size of cascades/avalanches, which are caused due to triggering of multiple such plastic modes at different spatial locations\cite{stanifer2022avalanche}. This is also consistent with the data in Fig.\ref{PressDens}(e) \& (f), 
where the magnitude of the drops in pressure/energy decrease with an increase in pinning concentration. The presence of a plasticity length scale in pinned solids also explains why there are negligible system size effects in the data shown in Fig. \ref{systemSize} as long as $\xi<<L$, which is the case for the system sizes that we report. As the length-scale $\xi$ prevents the system from acting as a whole beyond the lengths $l>\xi$.

\section{Summary \& Discussion}

To summarize, using extensive numerical simulations, we have investigated how  random inclusions, at the level of atoms/particles, influence the cavitation failure in amorphous solids. Our studies are done in the athermal quasistatic limit, i.e., where fluctuations originating from thermal fluctuations or finite deformation rates are absent. Like in previous studies, we model the inclusions as randomly pinned particles that only undergo affine motion during the mechanical deformation process but do not undergo any non-affine motion during the plastic relaxation.

We demonstrate how an amorphous solid with a small concentration of micro-inclusions, intended to make the solid stronger, can delay the cavitation and increase the tensile strength. The resulting glass obtained from the inclusions is thus not only stronger glass but also less susceptible to catastrophic fracture failure via cavitation. From a thermodynamic point of view, the delay in cavitation and, thus, the increase in tensile strength is likely a result of a change in the densities of coexisting solid and gas phases in the temperature-density ~\cite{sastry2000liquid,testard2011influence} phase diagram of the amorphous solid. A sharp yielding-like cavitation transition observed in amorphous solids ~\cite{Pablo,FirstWork} accompanied by an extensive peak in the pressure fluctuations characterized via a susceptibility $\chi_p$ is suppressed by particle-pinning. Particle pinning not only increases the tensile strength but also decreases the size of avalanches that lead to cavitation in amorphous solids by restricting them spatially; as a result, the cavitation occurs more homogeneously in this pinned solid. The expansion response of a pinned solid is somewhat analogous to the shear response of a ductile solid, as the sharp macroscopic drop leading to yielding is absent. But the analogy with a ductile solid doesn't hold completely as the yielding via cavitation is delayed in pinned solids. In a ductile solid, the yielding under shear would occur at lower strains and bulk stress ~\cite{BerthierCyclicShear}. Also, with increasing system sizes, the ductile solid shows brittle behaviour under shear via shear-banding-like events, as claimed in a recent work~\cite{richard2021finite}. In our range of probing, the finite size effects in the pressure $P$ vs $\rho$ curves and the pressure fluctuation $\chi_p$ vs $\rho$ curves do not show such a brittle-like character for the large system sizes. The absence of finite-size effects in the $P-\rho$ and $\chi_{p}-\rho$ curves, along with decreasing sizes of avalanches, is explained using a length-scale extracted from the spatial decay of displacements in plastic-eigenmodes.    

We have modeled these micro-alloyed inclusions via random pinning, which is a drastic simplifying approximation. Future studies can perhaps create more realistic models of inclusions in amorphous solids and study their deformation response. In Ref.~\cite{das2017pinning}, it was found that a particle that has a larger size compared to the typical size of the particles of the host medium can act as temporary pinning sites (termed as ``soft pinning'') over a timescale relevant to the medium relaxation time. Similarly, particles having different geometric shapes, like rods, can have significantly lower mobility due to their enhanced caging effect in a crowded environment and can also act as possible micro-alloying agents. Recent experimental realization of this ``soft pinning'' concept in colloidal glasses \cite{ganapathi2018measurements}, molecular glasses \cite{das2021soft}, and glassy polymer mixture \cite{kikumoto2020towards} is indeed very encouraging, indicating a clear possibility to test some of our results in experiments.

\section{Acknowledgements}

We thank the HPC facility at IMSc-Chennai for providing computational resources. S.K. would like to acknowledge the support from Swarna Jayanti Fellowship Grants No. DST/SJF/PSA-01/2018-19 and No.  SB/SFJ/2019-20/05 and Core Research Grant from SERB via grant CRG/2019/005373. PC acknowledges funding from SERB via grant MTR/2022/001034.

%\clearpage
\bibliography{References}

\begin{thebibliography}{51}%
\makeatletter
\providecommand \@ifxundefined [1]{%
 \@ifx{#1\undefined}
}%
\providecommand \@ifnum [1]{%
 \ifnum #1\expandafter \@firstoftwo
 \else \expandafter \@secondoftwo
 \fi
}%
\providecommand \@ifx [1]{%
 \ifx #1\expandafter \@firstoftwo
 \else \expandafter \@secondoftwo
 \fi
}%
\providecommand \natexlab [1]{#1}%
\providecommand \enquote  [1]{``#1''}%
\providecommand \bibnamefont  [1]{#1}%
\providecommand \bibfnamefont [1]{#1}%
\providecommand \citenamefont [1]{#1}%
\providecommand \href@noop [0]{\@secondoftwo}%
\providecommand \href [0]{\begingroup \@sanitize@url \@href}%
\providecommand \@href[1]{\@@startlink{#1}\@@href}%
\providecommand \@@href[1]{\endgroup#1\@@endlink}%
\providecommand \@sanitize@url [0]{\catcode `\\12\catcode `\$12\catcode
  `\&12\catcode `\#12\catcode `\^12\catcode `\_12\catcode `\%12\relax}%
\providecommand \@@startlink[1]{}%
\providecommand \@@endlink[0]{}%
\providecommand \url  [0]{\begingroup\@sanitize@url \@url }%
\providecommand \@url [1]{\endgroup\@href {#1}{\urlprefix }}%
\providecommand \urlprefix  [0]{URL }%
\providecommand \Eprint [0]{\href }%
\providecommand \doibase [0]{http://dx.doi.org/}%
\providecommand \selectlanguage [0]{\@gobble}%
\providecommand \bibinfo  [0]{\@secondoftwo}%
\providecommand \bibfield  [0]{\@secondoftwo}%
\providecommand \translation [1]{[#1]}%
\providecommand \BibitemOpen [0]{}%
\providecommand \bibitemStop [0]{}%
\providecommand \bibitemNoStop [0]{.\EOS\space}%
\providecommand \EOS [0]{\spacefactor3000\relax}%
\providecommand \BibitemShut  [1]{\csname bibitem#1\endcsname}%
\let\auto@bib@innerbib\@empty
%</preamble>
\bibitem [{\citenamefont {Schuh}\ \emph {et~al.}(2007)\citenamefont {Schuh},
  \citenamefont {Hufnagel},\ and\ \citenamefont
  {Ramamurty}}]{schuh2007mechanical}%
  \BibitemOpen
  \bibfield  {author} {\bibinfo {author} {\bibfnamefont {C.~A.}\ \bibnamefont
  {Schuh}}, \bibinfo {author} {\bibfnamefont {T.~C.}\ \bibnamefont {Hufnagel}},
  \ and\ \bibinfo {author} {\bibfnamefont {U.}~\bibnamefont {Ramamurty}},\
  }\href@noop {} {\bibfield  {journal} {\bibinfo  {journal} {Acta Materialia}\
  }\textbf {\bibinfo {volume} {55}},\ \bibinfo {pages} {4067} (\bibinfo {year}
  {2007})}\BibitemShut {NoStop}%
\bibitem [{\citenamefont {Bonn}\ \emph {et~al.}(2017)\citenamefont {Bonn},
  \citenamefont {Denn}, \citenamefont {Berthier}, \citenamefont {Divoux},\ and\
  \citenamefont {Manneville}}]{bonn2017yield}%
  \BibitemOpen
  \bibfield  {author} {\bibinfo {author} {\bibfnamefont {D.}~\bibnamefont
  {Bonn}}, \bibinfo {author} {\bibfnamefont {M.~M.}\ \bibnamefont {Denn}},
  \bibinfo {author} {\bibfnamefont {L.}~\bibnamefont {Berthier}}, \bibinfo
  {author} {\bibfnamefont {T.}~\bibnamefont {Divoux}}, \ and\ \bibinfo {author}
  {\bibfnamefont {S.}~\bibnamefont {Manneville}},\ }\href@noop {} {\bibfield
  {journal} {\bibinfo  {journal} {Reviews of Modern Physics}\ }\textbf
  {\bibinfo {volume} {89}},\ \bibinfo {pages} {035005} (\bibinfo {year}
  {2017})}\BibitemShut {NoStop}%
\bibitem [{\citenamefont {Rodney}\ \emph {et~al.}(2011)\citenamefont {Rodney},
  \citenamefont {Tanguy},\ and\ \citenamefont
  {Vandembroucq}}]{rodney2011modeling}%
  \BibitemOpen
  \bibfield  {author} {\bibinfo {author} {\bibfnamefont {D.}~\bibnamefont
  {Rodney}}, \bibinfo {author} {\bibfnamefont {A.}~\bibnamefont {Tanguy}}, \
  and\ \bibinfo {author} {\bibfnamefont {D.}~\bibnamefont {Vandembroucq}},\
  }\href@noop {} {\bibfield  {journal} {\bibinfo  {journal} {Modelling and
  Simulation in Materials Science and Engineering}\ }\textbf {\bibinfo {volume}
  {19}},\ \bibinfo {pages} {083001} (\bibinfo {year} {2011})}\BibitemShut
  {NoStop}%
\bibitem [{\citenamefont {Liu}\ \emph {et~al.}(2021)\citenamefont {Liu},
  \citenamefont {Dutta}, \citenamefont {Chaudhuri},\ and\ \citenamefont
  {Martens}}]{liu2021elastoplastic}%
  \BibitemOpen
  \bibfield  {author} {\bibinfo {author} {\bibfnamefont {C.}~\bibnamefont
  {Liu}}, \bibinfo {author} {\bibfnamefont {S.}~\bibnamefont {Dutta}}, \bibinfo
  {author} {\bibfnamefont {P.}~\bibnamefont {Chaudhuri}}, \ and\ \bibinfo
  {author} {\bibfnamefont {K.}~\bibnamefont {Martens}},\ }\href@noop {}
  {\bibfield  {journal} {\bibinfo  {journal} {Physical Review Letters}\
  }\textbf {\bibinfo {volume} {126}},\ \bibinfo {pages} {138005} (\bibinfo
  {year} {2021})}\BibitemShut {NoStop}%
\bibitem [{\citenamefont {Bouchaud}\ \emph {et~al.}(2008)\citenamefont
  {Bouchaud}, \citenamefont {Boivin}, \citenamefont {Pouchou}, \citenamefont
  {Bonamy}, \citenamefont {Poon},\ and\ \citenamefont
  {Ravichandran}}]{bouchaud2008fracture}%
  \BibitemOpen
  \bibfield  {author} {\bibinfo {author} {\bibfnamefont {E.}~\bibnamefont
  {Bouchaud}}, \bibinfo {author} {\bibfnamefont {D.}~\bibnamefont {Boivin}},
  \bibinfo {author} {\bibfnamefont {J.-L.}\ \bibnamefont {Pouchou}}, \bibinfo
  {author} {\bibfnamefont {D.}~\bibnamefont {Bonamy}}, \bibinfo {author}
  {\bibfnamefont {B.}~\bibnamefont {Poon}}, \ and\ \bibinfo {author}
  {\bibfnamefont {G.}~\bibnamefont {Ravichandran}},\ }\href@noop {} {\bibfield
  {journal} {\bibinfo  {journal} {Europhysics Letters}\ }\textbf {\bibinfo
  {volume} {83}},\ \bibinfo {pages} {66006} (\bibinfo {year}
  {2008})}\BibitemShut {NoStop}%
\bibitem [{\citenamefont {Murali}\ \emph {et~al.}(2011)\citenamefont {Murali},
  \citenamefont {Guo}, \citenamefont {Zhang}, \citenamefont {Narasimhan},
  \citenamefont {Li},\ and\ \citenamefont {Gao}}]{murali2011atomic}%
  \BibitemOpen
  \bibfield  {author} {\bibinfo {author} {\bibfnamefont {P.}~\bibnamefont
  {Murali}}, \bibinfo {author} {\bibfnamefont {T.}~\bibnamefont {Guo}},
  \bibinfo {author} {\bibfnamefont {Y.}~\bibnamefont {Zhang}}, \bibinfo
  {author} {\bibfnamefont {R.}~\bibnamefont {Narasimhan}}, \bibinfo {author}
  {\bibfnamefont {Y.}~\bibnamefont {Li}}, \ and\ \bibinfo {author}
  {\bibfnamefont {H.}~\bibnamefont {Gao}},\ }\href@noop {} {\bibfield
  {journal} {\bibinfo  {journal} {Physical Review Letters}\ }\textbf {\bibinfo
  {volume} {107}},\ \bibinfo {pages} {215501} (\bibinfo {year}
  {2011})}\BibitemShut {NoStop}%
\bibitem [{\citenamefont {Shen}\ \emph {et~al.}(2021)\citenamefont {Shen},
  \citenamefont {Yu}, \citenamefont {Tang}, \citenamefont {Sun}, \citenamefont
  {Liu}, \citenamefont {Bai},\ and\ \citenamefont
  {Wang}}]{shen2021observation}%
  \BibitemOpen
  \bibfield  {author} {\bibinfo {author} {\bibfnamefont {L.-Q.}\ \bibnamefont
  {Shen}}, \bibinfo {author} {\bibfnamefont {J.-H.}\ \bibnamefont {Yu}},
  \bibinfo {author} {\bibfnamefont {X.-C.}\ \bibnamefont {Tang}}, \bibinfo
  {author} {\bibfnamefont {B.-A.}\ \bibnamefont {Sun}}, \bibinfo {author}
  {\bibfnamefont {Y.-H.}\ \bibnamefont {Liu}}, \bibinfo {author} {\bibfnamefont
  {H.-Y.}\ \bibnamefont {Bai}}, \ and\ \bibinfo {author} {\bibfnamefont
  {W.-H.}\ \bibnamefont {Wang}},\ }\href@noop {} {\bibfield  {journal}
  {\bibinfo  {journal} {Science Advances}\ }\textbf {\bibinfo {volume} {7}},\
  \bibinfo {pages} {eabf7293} (\bibinfo {year} {2021})}\BibitemShut {NoStop}%
\bibitem [{\citenamefont {Guan}\ \emph {et~al.}(2013)\citenamefont {Guan},
  \citenamefont {Lu}, \citenamefont {Spector}, \citenamefont {Valavala},\ and\
  \citenamefont {Falk}}]{Falk}%
  \BibitemOpen
  \bibfield  {author} {\bibinfo {author} {\bibfnamefont {P.}~\bibnamefont
  {Guan}}, \bibinfo {author} {\bibfnamefont {S.}~\bibnamefont {Lu}}, \bibinfo
  {author} {\bibfnamefont {M.~J.~B.}\ \bibnamefont {Spector}}, \bibinfo
  {author} {\bibfnamefont {P.~K.}\ \bibnamefont {Valavala}}, \ and\ \bibinfo
  {author} {\bibfnamefont {M.~L.}\ \bibnamefont {Falk}},\ }\href {\doibase
  10.1103/PhysRevLett.110.185502} {\bibfield  {journal} {\bibinfo  {journal}
  {Phys. Rev. Lett.}\ }\textbf {\bibinfo {volume} {110}},\ \bibinfo {pages}
  {185502} (\bibinfo {year} {2013})}\BibitemShut {NoStop}%
\bibitem [{\citenamefont {Chaudhuri}\ and\ \citenamefont
  {Horbach}(2016)}]{Pinaki}%
  \BibitemOpen
  \bibfield  {author} {\bibinfo {author} {\bibfnamefont {P.}~\bibnamefont
  {Chaudhuri}}\ and\ \bibinfo {author} {\bibfnamefont {J.}~\bibnamefont
  {Horbach}},\ }\href@noop {} {\bibfield  {journal} {\bibinfo  {journal}
  {Physical Review B}\ }\textbf {\bibinfo {volume} {94}},\ \bibinfo {pages}
  {094203} (\bibinfo {year} {2016})}\BibitemShut {NoStop}%
\bibitem [{\citenamefont {Altabet}\ \emph
  {et~al.}(2018{\natexlab{a}})\citenamefont {Altabet}, \citenamefont {Fenley},
  \citenamefont {Stillinger},\ and\ \citenamefont {Debenedetti}}]{Pablo}%
  \BibitemOpen
  \bibfield  {author} {\bibinfo {author} {\bibfnamefont {Y.~E.}\ \bibnamefont
  {Altabet}}, \bibinfo {author} {\bibfnamefont {A.~L.}\ \bibnamefont {Fenley}},
  \bibinfo {author} {\bibfnamefont {F.~H.}\ \bibnamefont {Stillinger}}, \ and\
  \bibinfo {author} {\bibfnamefont {P.~G.}\ \bibnamefont {Debenedetti}},\
  }\href {\doibase 10.1063/1.5019274} {\bibfield  {journal} {\bibinfo
  {journal} {The Journal of Chemical Physics}\ }\textbf {\bibinfo {volume}
  {148}},\ \bibinfo {pages} {114501} (\bibinfo {year} {2018}{\natexlab{a}})},\
  \Eprint {http://arxiv.org/abs/https://doi.org/10.1063/1.5019274}
  {https://doi.org/10.1063/1.5019274} \BibitemShut {NoStop}%
\bibitem [{\citenamefont {Altabet}\ \emph
  {et~al.}(2018{\natexlab{b}})\citenamefont {Altabet}, \citenamefont {Fenley},
  \citenamefont {Stillinger},\ and\ \citenamefont {Debenedetti}}]{Pablo2}%
  \BibitemOpen
  \bibfield  {author} {\bibinfo {author} {\bibfnamefont {Y.~E.}\ \bibnamefont
  {Altabet}}, \bibinfo {author} {\bibfnamefont {A.~L.}\ \bibnamefont {Fenley}},
  \bibinfo {author} {\bibfnamefont {F.~H.}\ \bibnamefont {Stillinger}}, \ and\
  \bibinfo {author} {\bibfnamefont {P.~G.}\ \bibnamefont {Debenedetti}},\
  }\href {\doibase 10.1063/1.5019274} {\bibfield  {journal} {\bibinfo
  {journal} {The Journal of Chemical Physics}\ }\textbf {\bibinfo {volume}
  {148}},\ \bibinfo {pages} {114501} (\bibinfo {year}
  {2018}{\natexlab{b}})}\BibitemShut {NoStop}%
\bibitem [{\citenamefont {Paul}\ \emph {et~al.}(2020)\citenamefont {Paul},
  \citenamefont {Dasgupta}, \citenamefont {Horbach},\ and\ \citenamefont
  {Karmakar}}]{CavitySmarajit}%
  \BibitemOpen
  \bibfield  {author} {\bibinfo {author} {\bibfnamefont {K.}~\bibnamefont
  {Paul}}, \bibinfo {author} {\bibfnamefont {R.}~\bibnamefont {Dasgupta}},
  \bibinfo {author} {\bibfnamefont {J.}~\bibnamefont {Horbach}}, \ and\
  \bibinfo {author} {\bibfnamefont {S.}~\bibnamefont {Karmakar}},\ }\href
  {\doibase 10.1103/PhysRevResearch.2.042012} {\bibfield  {journal} {\bibinfo
  {journal} {Phys. Rev. Research}\ }\textbf {\bibinfo {volume} {2}},\ \bibinfo
  {pages} {042012} (\bibinfo {year} {2020})}\BibitemShut {NoStop}%
\bibitem [{\citenamefont {Dattani}\ \emph {et~al.}(2022)\citenamefont
  {Dattani}, \citenamefont {Karmakar},\ and\ \citenamefont
  {Chaudhuri}}]{FirstWork}%
  \BibitemOpen
  \bibfield  {author} {\bibinfo {author} {\bibfnamefont {U.~A.}\ \bibnamefont
  {Dattani}}, \bibinfo {author} {\bibfnamefont {S.}~\bibnamefont {Karmakar}}, \
  and\ \bibinfo {author} {\bibfnamefont {P.}~\bibnamefont {Chaudhuri}},\ }\href
  {\doibase 10.1103/PhysRevE.106.055004} {\bibfield  {journal} {\bibinfo
  {journal} {Phys. Rev. E}\ }\textbf {\bibinfo {volume} {106}},\ \bibinfo
  {pages} {055004} (\bibinfo {year} {2022})}\BibitemShut {NoStop}%
\bibitem [{\citenamefont {Dattani}\ \emph {et~al.}(2023)\citenamefont
  {Dattani}, \citenamefont {Sharma}, \citenamefont {Karmakar},\ and\
  \citenamefont {Chaudhuri}}]{SecondWork}%
  \BibitemOpen
  \bibfield  {author} {\bibinfo {author} {\bibfnamefont {U.~A.}\ \bibnamefont
  {Dattani}}, \bibinfo {author} {\bibfnamefont {R.}~\bibnamefont {Sharma}},
  \bibinfo {author} {\bibfnamefont {S.}~\bibnamefont {Karmakar}}, \ and\
  \bibinfo {author} {\bibfnamefont {P.}~\bibnamefont {Chaudhuri}},\ }\href@noop
  {} {\bibfield  {journal} {\bibinfo  {journal} {arXiv preprint
  arXiv:2303.04529}\ } (\bibinfo {year} {2023})}\BibitemShut {NoStop}%
\bibitem [{\citenamefont {Hofmann}\ \emph {et~al.}(2008)\citenamefont
  {Hofmann}, \citenamefont {Suh}, \citenamefont {Wiest}, \citenamefont {Duan},
  \citenamefont {Lind}, \citenamefont {Demetriou},\ and\ \citenamefont
  {Johnson}}]{PinningNature}%
  \BibitemOpen
  \bibfield  {author} {\bibinfo {author} {\bibfnamefont {D.~C.}\ \bibnamefont
  {Hofmann}}, \bibinfo {author} {\bibfnamefont {J.-Y.}\ \bibnamefont {Suh}},
  \bibinfo {author} {\bibfnamefont {A.}~\bibnamefont {Wiest}}, \bibinfo
  {author} {\bibfnamefont {G.}~\bibnamefont {Duan}}, \bibinfo {author}
  {\bibfnamefont {M.-L.}\ \bibnamefont {Lind}}, \bibinfo {author}
  {\bibfnamefont {M.~D.}\ \bibnamefont {Demetriou}}, \ and\ \bibinfo {author}
  {\bibfnamefont {W.~L.}\ \bibnamefont {Johnson}},\ }\href@noop {} {\bibfield
  {journal} {\bibinfo  {journal} {Nature}\ }\textbf {\bibinfo {volume} {451}},\
  \bibinfo {pages} {1085} (\bibinfo {year} {2008})}\BibitemShut {NoStop}%
\bibitem [{\citenamefont {Hardin}\ and\ \citenamefont
  {Homer}(2015)}]{PinningIntro}%
  \BibitemOpen
  \bibfield  {author} {\bibinfo {author} {\bibfnamefont {T.~J.}\ \bibnamefont
  {Hardin}}\ and\ \bibinfo {author} {\bibfnamefont {E.~R.}\ \bibnamefont
  {Homer}},\ }\href@noop {} {\bibfield  {journal} {\bibinfo  {journal} {Acta
  Materialia}\ }\textbf {\bibinfo {volume} {83}},\ \bibinfo {pages} {203}
  (\bibinfo {year} {2015})}\BibitemShut {NoStop}%
\bibitem [{\citenamefont {Torquato}\ and\ \citenamefont
  {Haslach~Jr}(2002)}]{torquato2002random}%
  \BibitemOpen
  \bibfield  {author} {\bibinfo {author} {\bibfnamefont {S.}~\bibnamefont
  {Torquato}}\ and\ \bibinfo {author} {\bibfnamefont {H.}~\bibnamefont
  {Haslach~Jr}},\ }\href@noop {} {\bibfield  {journal} {\bibinfo  {journal}
  {Appl. Mech. Rev.}\ }\textbf {\bibinfo {volume} {55}},\ \bibinfo {pages}
  {B62} (\bibinfo {year} {2002})}\BibitemShut {NoStop}%
\bibitem [{\citenamefont {Tyukodi}\ \emph {et~al.}(2016)\citenamefont
  {Tyukodi}, \citenamefont {Lemarchand}, \citenamefont {Hansen},\ and\
  \citenamefont {Vandembroucq}}]{tyukodi2016finite}%
  \BibitemOpen
  \bibfield  {author} {\bibinfo {author} {\bibfnamefont {B.}~\bibnamefont
  {Tyukodi}}, \bibinfo {author} {\bibfnamefont {C.~A.}\ \bibnamefont
  {Lemarchand}}, \bibinfo {author} {\bibfnamefont {J.~S.}\ \bibnamefont
  {Hansen}}, \ and\ \bibinfo {author} {\bibfnamefont {D.}~\bibnamefont
  {Vandembroucq}},\ }\href@noop {} {\bibfield  {journal} {\bibinfo  {journal}
  {Physical Review E}\ }\textbf {\bibinfo {volume} {93}},\ \bibinfo {pages}
  {023004} (\bibinfo {year} {2016})}\BibitemShut {NoStop}%
\bibitem [{\citenamefont {Garrett}\ \emph {et~al.}(2012)\citenamefont
  {Garrett}, \citenamefont {Demetriou}, \citenamefont {Chen},\ and\
  \citenamefont {Johnson}}]{garrett2012effect}%
  \BibitemOpen
  \bibfield  {author} {\bibinfo {author} {\bibfnamefont {G.~R.}\ \bibnamefont
  {Garrett}}, \bibinfo {author} {\bibfnamefont {M.~D.}\ \bibnamefont
  {Demetriou}}, \bibinfo {author} {\bibfnamefont {J.}~\bibnamefont {Chen}}, \
  and\ \bibinfo {author} {\bibfnamefont {W.~L.}\ \bibnamefont {Johnson}},\
  }\href@noop {} {\bibfield  {journal} {\bibinfo  {journal} {Applied Physics
  Letters}\ }\textbf {\bibinfo {volume} {101}},\ \bibinfo {pages} {241913}
  (\bibinfo {year} {2012})}\BibitemShut {NoStop}%
\bibitem [{\citenamefont {Gonz{\'a}lez}(2016)}]{gonzalez2016role}%
  \BibitemOpen
  \bibfield  {author} {\bibinfo {author} {\bibfnamefont {S.}~\bibnamefont
  {Gonz{\'a}lez}},\ }\href@noop {} {\bibfield  {journal} {\bibinfo  {journal}
  {Journal of Materials Research}\ }\textbf {\bibinfo {volume} {31}},\ \bibinfo
  {pages} {76} (\bibinfo {year} {2016})}\BibitemShut {NoStop}%
\bibitem [{\citenamefont {Gendelman}\ \emph {et~al.}(2014)\citenamefont
  {Gendelman}, \citenamefont {Joy}, \citenamefont {Mishra}, \citenamefont
  {Procaccia},\ and\ \citenamefont {Samwer}}]{ItamarPinning}%
  \BibitemOpen
  \bibfield  {author} {\bibinfo {author} {\bibfnamefont {O.}~\bibnamefont
  {Gendelman}}, \bibinfo {author} {\bibfnamefont {A.}~\bibnamefont {Joy}},
  \bibinfo {author} {\bibfnamefont {P.}~\bibnamefont {Mishra}}, \bibinfo
  {author} {\bibfnamefont {I.}~\bibnamefont {Procaccia}}, \ and\ \bibinfo
  {author} {\bibfnamefont {K.}~\bibnamefont {Samwer}},\ }\href@noop {}
  {\bibfield  {journal} {\bibinfo  {journal} {Acta materialia}\ }\textbf
  {\bibinfo {volume} {63}},\ \bibinfo {pages} {209} (\bibinfo {year}
  {2014})}\BibitemShut {NoStop}%
\bibitem [{\citenamefont {Bhowmik}\ \emph
  {et~al.}(2019{\natexlab{a}})\citenamefont {Bhowmik}, \citenamefont
  {Chaudhuri},\ and\ \citenamefont {Karmakar}}]{Bhanu}%
  \BibitemOpen
  \bibfield  {author} {\bibinfo {author} {\bibfnamefont {B.~P.}\ \bibnamefont
  {Bhowmik}}, \bibinfo {author} {\bibfnamefont {P.}~\bibnamefont {Chaudhuri}},
  \ and\ \bibinfo {author} {\bibfnamefont {S.}~\bibnamefont {Karmakar}},\
  }\href {\doibase 10.1103/PhysRevLett.123.185501} {\bibfield  {journal}
  {\bibinfo  {journal} {Phys. Rev. Lett.}\ }\textbf {\bibinfo {volume} {123}},\
  \bibinfo {pages} {185501} (\bibinfo {year} {2019}{\natexlab{a}})}\BibitemShut
  {NoStop}%
\bibitem [{\citenamefont {Bhowmik}\ \emph
  {et~al.}(2019{\natexlab{b}})\citenamefont {Bhowmik}, \citenamefont
  {Karmakar}, \citenamefont {Procaccia},\ and\ \citenamefont
  {Rainone}}]{Bhanu2}%
  \BibitemOpen
  \bibfield  {author} {\bibinfo {author} {\bibfnamefont {B.~P.}\ \bibnamefont
  {Bhowmik}}, \bibinfo {author} {\bibfnamefont {S.}~\bibnamefont {Karmakar}},
  \bibinfo {author} {\bibfnamefont {I.}~\bibnamefont {Procaccia}}, \ and\
  \bibinfo {author} {\bibfnamefont {C.}~\bibnamefont {Rainone}},\ }\href
  {\doibase 10.1103/PhysRevE.100.052110} {\bibfield  {journal} {\bibinfo
  {journal} {Phys. Rev. E}\ }\textbf {\bibinfo {volume} {100}},\ \bibinfo
  {pages} {052110} (\bibinfo {year} {2019}{\natexlab{b}})}\BibitemShut
  {NoStop}%
\bibitem [{\citenamefont {Liu}\ \emph {et~al.}(2023)\citenamefont {Liu},
  \citenamefont {Liu},\ and\ \citenamefont {Peng}}]{liu2023pinningLengthScale}%
  \BibitemOpen
  \bibfield  {author} {\bibinfo {author} {\bibfnamefont {Y.}~\bibnamefont
  {Liu}}, \bibinfo {author} {\bibfnamefont {H.}~\bibnamefont {Liu}}, \ and\
  \bibinfo {author} {\bibfnamefont {H.}~\bibnamefont {Peng}},\ }\href@noop {}
  {\bibfield  {journal} {\bibinfo  {journal} {Journal of Non-Crystalline
  Solids}\ }\textbf {\bibinfo {volume} {601}},\ \bibinfo {pages} {122052}
  (\bibinfo {year} {2023})}\BibitemShut {NoStop}%
\bibitem [{\citenamefont {Brüning}\ \emph {et~al.}(2008)\citenamefont
  {Brüning}, \citenamefont {St-Onge}, \citenamefont {Patterson},\ and\
  \citenamefont {Kob}}]{2DKA}%
  \BibitemOpen
  \bibfield  {author} {\bibinfo {author} {\bibfnamefont {R.}~\bibnamefont
  {Brüning}}, \bibinfo {author} {\bibfnamefont {D.~A.}\ \bibnamefont
  {St-Onge}}, \bibinfo {author} {\bibfnamefont {S.}~\bibnamefont {Patterson}},
  \ and\ \bibinfo {author} {\bibfnamefont {W.}~\bibnamefont {Kob}},\ }\href
  {\doibase 10.1088/0953-8984/21/3/035117} {\bibfield  {journal} {\bibinfo
  {journal} {Journal of Physics: Condensed Matter}\ }\textbf {\bibinfo {volume}
  {21}},\ \bibinfo {pages} {035117} (\bibinfo {year} {2008})}\BibitemShut
  {NoStop}%
\bibitem [{\citenamefont {Polak}\ and\ \citenamefont {Ribiere}(1969)}]{cg}%
  \BibitemOpen
  \bibfield  {author} {\bibinfo {author} {\bibfnamefont {E.}~\bibnamefont
  {Polak}}\ and\ \bibinfo {author} {\bibfnamefont {G.}~\bibnamefont
  {Ribiere}},\ }\href@noop {} {\bibfield  {journal} {\bibinfo  {journal} {Revue
  fran{\c{c}}aise d'informatique et de recherche op{\'e}rationnelle. S{\'e}rie
  rouge}\ }\textbf {\bibinfo {volume} {3}},\ \bibinfo {pages} {35} (\bibinfo
  {year} {1969})}\BibitemShut {NoStop}%
\bibitem [{\citenamefont {Plimpton}(1995)}]{LAMMPS}%
  \BibitemOpen
  \bibfield  {author} {\bibinfo {author} {\bibfnamefont {S.}~\bibnamefont
  {Plimpton}},\ }\href@noop {} {\bibfield  {journal} {\bibinfo  {journal}
  {Journal of Computational Physics}\ }\textbf {\bibinfo {volume} {117}},\
  \bibinfo {pages} {1} (\bibinfo {year} {1995})}\BibitemShut {NoStop}%
\bibitem [{\citenamefont {Anderson}\ \emph {et~al.}(1999)\citenamefont
  {Anderson}, \citenamefont {Bai}, \citenamefont {Bischof}, \citenamefont
  {Blackford}, \citenamefont {Demmel}, \citenamefont {Dongarra}, \citenamefont
  {Du~Croz}, \citenamefont {Greenbaum}, \citenamefont {Hammarling},
  \citenamefont {McKenney},\ and\ \citenamefont {Sorensen}}]{lapack99}%
  \BibitemOpen
  \bibfield  {author} {\bibinfo {author} {\bibfnamefont {E.}~\bibnamefont
  {Anderson}}, \bibinfo {author} {\bibfnamefont {Z.}~\bibnamefont {Bai}},
  \bibinfo {author} {\bibfnamefont {C.}~\bibnamefont {Bischof}}, \bibinfo
  {author} {\bibfnamefont {S.}~\bibnamefont {Blackford}}, \bibinfo {author}
  {\bibfnamefont {J.}~\bibnamefont {Demmel}}, \bibinfo {author} {\bibfnamefont
  {J.}~\bibnamefont {Dongarra}}, \bibinfo {author} {\bibfnamefont
  {J.}~\bibnamefont {Du~Croz}}, \bibinfo {author} {\bibfnamefont
  {A.}~\bibnamefont {Greenbaum}}, \bibinfo {author} {\bibfnamefont
  {S.}~\bibnamefont {Hammarling}}, \bibinfo {author} {\bibfnamefont
  {A.}~\bibnamefont {McKenney}}, \ and\ \bibinfo {author} {\bibfnamefont
  {D.}~\bibnamefont {Sorensen}},\ }\href@noop {} {\emph {\bibinfo {title}
  {{LAPACK} Users' Guide}}},\ \bibinfo {edition} {3rd}\ ed.\ (\bibinfo
  {publisher} {Society for Industrial and Applied Mathematics},\ \bibinfo
  {address} {Philadelphia, PA},\ \bibinfo {year} {1999})\BibitemShut {NoStop}%
\bibitem [{\citenamefont {Malandro}\ and\ \citenamefont
  {Lacks}(1999)}]{MalandroLacks}%
  \BibitemOpen
  \bibfield  {author} {\bibinfo {author} {\bibfnamefont {D.~L.}\ \bibnamefont
  {Malandro}}\ and\ \bibinfo {author} {\bibfnamefont {D.~J.}\ \bibnamefont
  {Lacks}},\ }\href {\doibase 10.1063/1.478340} {\bibfield  {journal} {\bibinfo
   {journal} {The Journal of Chemical Physics}\ }\textbf {\bibinfo {volume}
  {110}},\ \bibinfo {pages} {4593} (\bibinfo {year} {1999})},\ \Eprint
  {http://arxiv.org/abs/https://doi.org/10.1063/1.478340}
  {https://doi.org/10.1063/1.478340} \BibitemShut {NoStop}%
\bibitem [{\citenamefont {Maloney}\ and\ \citenamefont
  {Lema\^{\i}tre}(2006)}]{MaloneyLemaitre}%
  \BibitemOpen
  \bibfield  {author} {\bibinfo {author} {\bibfnamefont {C.~E.}\ \bibnamefont
  {Maloney}}\ and\ \bibinfo {author} {\bibfnamefont {A.}~\bibnamefont
  {Lema\^{\i}tre}},\ }\href {\doibase 10.1103/PhysRevE.74.016118} {\bibfield
  {journal} {\bibinfo  {journal} {Phys. Rev. E}\ }\textbf {\bibinfo {volume}
  {74}},\ \bibinfo {pages} {016118} (\bibinfo {year} {2006})}\BibitemShut
  {NoStop}%
\bibitem [{\citenamefont {Karmakar}\ \emph
  {et~al.}(2010{\natexlab{a}})\citenamefont {Karmakar}, \citenamefont
  {Lerner},\ and\ \citenamefont {Procaccia}}]{SquareRootDerivation}%
  \BibitemOpen
  \bibfield  {author} {\bibinfo {author} {\bibfnamefont {S.}~\bibnamefont
  {Karmakar}}, \bibinfo {author} {\bibfnamefont {E.}~\bibnamefont {Lerner}}, \
  and\ \bibinfo {author} {\bibfnamefont {I.}~\bibnamefont {Procaccia}},\ }\href
  {\doibase 10.1103/PhysRevE.82.026105} {\bibfield  {journal} {\bibinfo
  {journal} {Phys. Rev. E}\ }\textbf {\bibinfo {volume} {82}},\ \bibinfo
  {pages} {026105} (\bibinfo {year} {2010}{\natexlab{a}})}\BibitemShut
  {NoStop}%
\bibitem [{\citenamefont {Lema\^{\i}tre}\ and\ \citenamefont
  {Caroli}(2009)}]{AValanches1}%
  \BibitemOpen
  \bibfield  {author} {\bibinfo {author} {\bibfnamefont {A.}~\bibnamefont
  {Lema\^{\i}tre}}\ and\ \bibinfo {author} {\bibfnamefont {C.}~\bibnamefont
  {Caroli}},\ }\href {\doibase 10.1103/PhysRevLett.103.065501} {\bibfield
  {journal} {\bibinfo  {journal} {Phys. Rev. Lett.}\ }\textbf {\bibinfo
  {volume} {103}},\ \bibinfo {pages} {065501} (\bibinfo {year}
  {2009})}\BibitemShut {NoStop}%
\bibitem [{\citenamefont {Karmakar}\ \emph
  {et~al.}(2010{\natexlab{b}})\citenamefont {Karmakar}, \citenamefont {Lerner},
  \citenamefont {Procaccia},\ and\ \citenamefont {Zylberg}}]{Avalanches2}%
  \BibitemOpen
  \bibfield  {author} {\bibinfo {author} {\bibfnamefont {S.}~\bibnamefont
  {Karmakar}}, \bibinfo {author} {\bibfnamefont {E.}~\bibnamefont {Lerner}},
  \bibinfo {author} {\bibfnamefont {I.}~\bibnamefont {Procaccia}}, \ and\
  \bibinfo {author} {\bibfnamefont {J.}~\bibnamefont {Zylberg}},\ }\href
  {\doibase 10.1103/PhysRevE.82.031301} {\bibfield  {journal} {\bibinfo
  {journal} {Phys. Rev. E}\ }\textbf {\bibinfo {volume} {82}},\ \bibinfo
  {pages} {031301} (\bibinfo {year} {2010}{\natexlab{b}})}\BibitemShut
  {NoStop}%
\bibitem [{\citenamefont {Niblett}\ \emph {et~al.}(2018)\citenamefont
  {Niblett}, \citenamefont {de~Souza}, \citenamefont {Jack},\ and\
  \citenamefont {Wales}}]{WalesEnergyLandscape}%
  \BibitemOpen
  \bibfield  {author} {\bibinfo {author} {\bibfnamefont {S.~P.}\ \bibnamefont
  {Niblett}}, \bibinfo {author} {\bibfnamefont {V.~K.}\ \bibnamefont
  {de~Souza}}, \bibinfo {author} {\bibfnamefont {R.~L.}\ \bibnamefont {Jack}},
  \ and\ \bibinfo {author} {\bibfnamefont {D.~J.}\ \bibnamefont {Wales}},\
  }\href {\doibase 10.1063/1.5042140} {\bibfield  {journal} {\bibinfo
  {journal} {The Journal of Chemical Physics}\ }\textbf {\bibinfo {volume}
  {149}},\ \bibinfo {pages} {114503} (\bibinfo {year} {2018})},\ \Eprint
  {http://arxiv.org/abs/https://doi.org/10.1063/1.5042140}
  {https://doi.org/10.1063/1.5042140} \BibitemShut {NoStop}%
\bibitem [{\citenamefont {Argon}\ and\ \citenamefont
  {Kuo}(1979)}]{Argon1979STZ}%
  \BibitemOpen
  \bibfield  {author} {\bibinfo {author} {\bibfnamefont {A.}~\bibnamefont
  {Argon}}\ and\ \bibinfo {author} {\bibfnamefont {H.}~\bibnamefont {Kuo}},\
  }\href {\doibase https://doi.org/10.1016/0025-5416(79)90174-5} {\bibfield
  {journal} {\bibinfo  {journal} {Materials Science and Engineering}\ }\textbf
  {\bibinfo {volume} {39}},\ \bibinfo {pages} {101} (\bibinfo {year}
  {1979})}\BibitemShut {NoStop}%
\bibitem [{\citenamefont {Falk}\ and\ \citenamefont
  {Langer}(1998)}]{FalkLangerSTZ}%
  \BibitemOpen
  \bibfield  {author} {\bibinfo {author} {\bibfnamefont {M.~L.}\ \bibnamefont
  {Falk}}\ and\ \bibinfo {author} {\bibfnamefont {J.~S.}\ \bibnamefont
  {Langer}},\ }\href {\doibase 10.1103/PhysRevE.57.7192} {\bibfield  {journal}
  {\bibinfo  {journal} {Phys. Rev. E}\ }\textbf {\bibinfo {volume} {57}},\
  \bibinfo {pages} {7192} (\bibinfo {year} {1998})}\BibitemShut {NoStop}%
\bibitem [{\citenamefont {Richard}\ \emph
  {et~al.}(2021{\natexlab{a}})\citenamefont {Richard}, \citenamefont
  {Kapteijns}, \citenamefont {Giannini}, \citenamefont {Manning},\ and\
  \citenamefont {Lerner}}]{LernerSTZ}%
  \BibitemOpen
  \bibfield  {author} {\bibinfo {author} {\bibfnamefont {D.}~\bibnamefont
  {Richard}}, \bibinfo {author} {\bibfnamefont {G.}~\bibnamefont {Kapteijns}},
  \bibinfo {author} {\bibfnamefont {J.~A.}\ \bibnamefont {Giannini}}, \bibinfo
  {author} {\bibfnamefont {M.~L.}\ \bibnamefont {Manning}}, \ and\ \bibinfo
  {author} {\bibfnamefont {E.}~\bibnamefont {Lerner}},\ }\href {\doibase
  10.1103/PhysRevLett.126.015501} {\bibfield  {journal} {\bibinfo  {journal}
  {Phys. Rev. Lett.}\ }\textbf {\bibinfo {volume} {126}},\ \bibinfo {pages}
  {015501} (\bibinfo {year} {2021}{\natexlab{a}})}\BibitemShut {NoStop}%
\bibitem [{\citenamefont {Eshelby}(1957)}]{eshelby1957determination}%
  \BibitemOpen
  \bibfield  {author} {\bibinfo {author} {\bibfnamefont {J.~D.}\ \bibnamefont
  {Eshelby}},\ }\href@noop {} {\bibfield  {journal} {\bibinfo  {journal}
  {Proceedings of the royal society of London. Series A. Mathematical and
  physical sciences}\ }\textbf {\bibinfo {volume} {241}},\ \bibinfo {pages}
  {376} (\bibinfo {year} {1957})}\BibitemShut {NoStop}%
\bibitem [{\citenamefont {Gartner}\ and\ \citenamefont
  {Lerner}(2016)}]{LukaLernerPlasticModes}%
  \BibitemOpen
  \bibfield  {author} {\bibinfo {author} {\bibfnamefont {L.}~\bibnamefont
  {Gartner}}\ and\ \bibinfo {author} {\bibfnamefont {E.}~\bibnamefont
  {Lerner}},\ }\href {\doibase 10.1103/PhysRevE.93.011001} {\bibfield
  {journal} {\bibinfo  {journal} {Phys. Rev. E}\ }\textbf {\bibinfo {volume}
  {93}},\ \bibinfo {pages} {011001} (\bibinfo {year} {2016})}\BibitemShut
  {NoStop}%
\bibitem [{\citenamefont {Tanguy}\ \emph {et~al.}(2010)\citenamefont {Tanguy},
  \citenamefont {Mantisi},\ and\ \citenamefont {Tsamados}}]{EigenmodesTanguy}%
  \BibitemOpen
  \bibfield  {author} {\bibinfo {author} {\bibfnamefont {A.}~\bibnamefont
  {Tanguy}}, \bibinfo {author} {\bibfnamefont {B.}~\bibnamefont {Mantisi}}, \
  and\ \bibinfo {author} {\bibfnamefont {M.}~\bibnamefont {Tsamados}},\ }\href
  {\doibase 10.1209/0295-5075/90/16004} {\bibfield  {journal} {\bibinfo
  {journal} {Europhysics Letters}\ }\textbf {\bibinfo {volume} {90}},\ \bibinfo
  {pages} {16004} (\bibinfo {year} {2010})}\BibitemShut {NoStop}%
\bibitem [{\citenamefont {Manning}\ and\ \citenamefont
  {Liu}(2011)}]{SoftModesManningLiu}%
  \BibitemOpen
  \bibfield  {author} {\bibinfo {author} {\bibfnamefont {M.~L.}\ \bibnamefont
  {Manning}}\ and\ \bibinfo {author} {\bibfnamefont {A.~J.}\ \bibnamefont
  {Liu}},\ }\href {\doibase 10.1103/PhysRevLett.107.108302} {\bibfield
  {journal} {\bibinfo  {journal} {Phys. Rev. Lett.}\ }\textbf {\bibinfo
  {volume} {107}},\ \bibinfo {pages} {108302} (\bibinfo {year}
  {2011})}\BibitemShut {NoStop}%
\bibitem [{Note1()}]{Note1}%
  \BibitemOpen
  \bibinfo {note} {Defined as particle undergoing highest displacement in an
  eigenmode}\BibitemShut {NoStop}%
\bibitem [{\citenamefont {Stanifer}\ and\ \citenamefont
  {Manning}(2022)}]{stanifer2022avalanche}%
  \BibitemOpen
  \bibfield  {author} {\bibinfo {author} {\bibfnamefont {E.}~\bibnamefont
  {Stanifer}}\ and\ \bibinfo {author} {\bibfnamefont {M.~L.}\ \bibnamefont
  {Manning}},\ }\href@noop {} {\bibfield  {journal} {\bibinfo  {journal} {Soft
  Matter}\ }\textbf {\bibinfo {volume} {18}},\ \bibinfo {pages} {2394}
  (\bibinfo {year} {2022})}\BibitemShut {NoStop}%
\bibitem [{\citenamefont {Sastry}(2000)}]{sastry2000liquid}%
  \BibitemOpen
  \bibfield  {author} {\bibinfo {author} {\bibfnamefont {S.}~\bibnamefont
  {Sastry}},\ }\href@noop {} {\bibfield  {journal} {\bibinfo  {journal}
  {Physical Review Letters}\ }\textbf {\bibinfo {volume} {85}},\ \bibinfo
  {pages} {590} (\bibinfo {year} {2000})}\BibitemShut {NoStop}%
\bibitem [{\citenamefont {Testard}\ \emph {et~al.}(2011)\citenamefont
  {Testard}, \citenamefont {Berthier},\ and\ \citenamefont
  {Kob}}]{testard2011influence}%
  \BibitemOpen
  \bibfield  {author} {\bibinfo {author} {\bibfnamefont {V.}~\bibnamefont
  {Testard}}, \bibinfo {author} {\bibfnamefont {L.}~\bibnamefont {Berthier}}, \
  and\ \bibinfo {author} {\bibfnamefont {W.}~\bibnamefont {Kob}},\ }\href@noop
  {} {\bibfield  {journal} {\bibinfo  {journal} {Physical review letters}\
  }\textbf {\bibinfo {volume} {106}},\ \bibinfo {pages} {125702} (\bibinfo
  {year} {2011})}\BibitemShut {NoStop}%
\bibitem [{\citenamefont {Yeh}\ \emph {et~al.}(2020)\citenamefont {Yeh},
  \citenamefont {Ozawa}, \citenamefont {Miyazaki}, \citenamefont {Kawasaki},\
  and\ \citenamefont {Berthier}}]{BerthierCyclicShear}%
  \BibitemOpen
  \bibfield  {author} {\bibinfo {author} {\bibfnamefont {W.-T.}\ \bibnamefont
  {Yeh}}, \bibinfo {author} {\bibfnamefont {M.}~\bibnamefont {Ozawa}}, \bibinfo
  {author} {\bibfnamefont {K.}~\bibnamefont {Miyazaki}}, \bibinfo {author}
  {\bibfnamefont {T.}~\bibnamefont {Kawasaki}}, \ and\ \bibinfo {author}
  {\bibfnamefont {L.}~\bibnamefont {Berthier}},\ }\href {\doibase
  10.1103/PhysRevLett.124.225502} {\bibfield  {journal} {\bibinfo  {journal}
  {Phys. Rev. Lett.}\ }\textbf {\bibinfo {volume} {124}},\ \bibinfo {pages}
  {225502} (\bibinfo {year} {2020})}\BibitemShut {NoStop}%
\bibitem [{\citenamefont {Richard}\ \emph
  {et~al.}(2021{\natexlab{b}})\citenamefont {Richard}, \citenamefont
  {Rainone},\ and\ \citenamefont {Lerner}}]{richard2021finite}%
  \BibitemOpen
  \bibfield  {author} {\bibinfo {author} {\bibfnamefont {D.}~\bibnamefont
  {Richard}}, \bibinfo {author} {\bibfnamefont {C.}~\bibnamefont {Rainone}}, \
  and\ \bibinfo {author} {\bibfnamefont {E.}~\bibnamefont {Lerner}},\
  }\href@noop {} {\bibfield  {journal} {\bibinfo  {journal} {The Journal of
  Chemical Physics}\ }\textbf {\bibinfo {volume} {155}},\ \bibinfo {pages}
  {056101} (\bibinfo {year} {2021}{\natexlab{b}})}\BibitemShut {NoStop}%
\bibitem [{\citenamefont {Das}\ \emph {et~al.}(2017)\citenamefont {Das},
  \citenamefont {Chakrabarty},\ and\ \citenamefont
  {Karmakar}}]{das2017pinning}%
  \BibitemOpen
  \bibfield  {author} {\bibinfo {author} {\bibfnamefont {R.}~\bibnamefont
  {Das}}, \bibinfo {author} {\bibfnamefont {S.}~\bibnamefont {Chakrabarty}}, \
  and\ \bibinfo {author} {\bibfnamefont {S.}~\bibnamefont {Karmakar}},\
  }\href@noop {} {\bibfield  {journal} {\bibinfo  {journal} {Soft matter}\
  }\textbf {\bibinfo {volume} {13}},\ \bibinfo {pages} {6929} (\bibinfo {year}
  {2017})}\BibitemShut {NoStop}%
\bibitem [{\citenamefont {Ganapathi}\ \emph {et~al.}(2018)\citenamefont
  {Ganapathi}, \citenamefont {Nagamanasa}, \citenamefont {Sood},\ and\
  \citenamefont {Ganapathy}}]{ganapathi2018measurements}%
  \BibitemOpen
  \bibfield  {author} {\bibinfo {author} {\bibfnamefont {D.}~\bibnamefont
  {Ganapathi}}, \bibinfo {author} {\bibfnamefont {K.~H.}\ \bibnamefont
  {Nagamanasa}}, \bibinfo {author} {\bibfnamefont {A.}~\bibnamefont {Sood}}, \
  and\ \bibinfo {author} {\bibfnamefont {R.}~\bibnamefont {Ganapathy}},\
  }\href@noop {} {\bibfield  {journal} {\bibinfo  {journal} {Nature
  communications}\ }\textbf {\bibinfo {volume} {9}},\ \bibinfo {pages} {397}
  (\bibinfo {year} {2018})}\BibitemShut {NoStop}%
\bibitem [{\citenamefont {Das}\ \emph {et~al.}(2021)\citenamefont {Das},
  \citenamefont {Bhowmik}, \citenamefont {Puthirath}, \citenamefont
  {Narayanan},\ and\ \citenamefont {Karmakar}}]{das2021soft}%
  \BibitemOpen
  \bibfield  {author} {\bibinfo {author} {\bibfnamefont {R.}~\bibnamefont
  {Das}}, \bibinfo {author} {\bibfnamefont {B.~P.}\ \bibnamefont {Bhowmik}},
  \bibinfo {author} {\bibfnamefont {A.~B.}\ \bibnamefont {Puthirath}}, \bibinfo
  {author} {\bibfnamefont {T.~N.}\ \bibnamefont {Narayanan}}, \ and\ \bibinfo
  {author} {\bibfnamefont {S.}~\bibnamefont {Karmakar}},\ }\href@noop {}
  {\bibfield  {journal} {\bibinfo  {journal} {arXiv preprint arXiv:2106.06325}\
  } (\bibinfo {year} {2021})}\BibitemShut {NoStop}%
\bibitem [{\citenamefont {Kikumoto}\ \emph {et~al.}(2020)\citenamefont
  {Kikumoto}, \citenamefont {Torii}, \citenamefont {Fukao}, \citenamefont
  {Royall}, \citenamefont {Yao}, \citenamefont {Saruyama},\ and\ \citenamefont
  {Tatsumi}}]{kikumoto2020towards}%
  \BibitemOpen
  \bibfield  {author} {\bibinfo {author} {\bibfnamefont {G.}~\bibnamefont
  {Kikumoto}}, \bibinfo {author} {\bibfnamefont {N.}~\bibnamefont {Torii}},
  \bibinfo {author} {\bibfnamefont {K.}~\bibnamefont {Fukao}}, \bibinfo
  {author} {\bibfnamefont {C.~P.}\ \bibnamefont {Royall}}, \bibinfo {author}
  {\bibfnamefont {H.}~\bibnamefont {Yao}}, \bibinfo {author} {\bibfnamefont
  {Y.}~\bibnamefont {Saruyama}}, \ and\ \bibinfo {author} {\bibfnamefont
  {S.}~\bibnamefont {Tatsumi}},\ }\href@noop {} {\bibfield  {journal} {\bibinfo
   {journal} {arXiv preprint arXiv:2003.06089}\ } (\bibinfo {year}
  {2020})}\BibitemShut {NoStop}%
\end{thebibliography}%

\end{document}